\documentclass[12pt]{article}%
\usepackage{amsmath,amssymb,amsthm}%
\newtheorem{theorem}{Theorem}
\begin{document}

\title{Extensors\thanks{published: \emph{Advances in Applied Clifford Algebras}
\textbf{11}(S3), 23-40 (2001).}}
\author{Virginia V. Fern\'{a}ndez$^{1}\thanks{e-mail: vvf@ime.unicamp.br}$, Antonio M.
Moya$^1\thanks{e-mail: moya@ime.unicamp.br}$
\and and Waldyr A. Rodrigues Jr.$^{1,2}\thanks{e-mail: walrod@ime.unicamp.br or
walrod@mpc.com.br}$\\$^{1}$Institute of Mathematics, Statistics and Scientific Computation \\IMECC-UNICAMP CP 6065\\13083-970 Campinas-SP, Brazil\\$^{2}$Department of Mathematical Sciences, University\ of Liverpool\\Liverpool L69 3BX, UK}
\date{11/26/2001}
\maketitle

\begin{abstract}
In this paper we introduce a class of mathematical objects called
\emph{extensors} and develop some aspects of their theory with considerable
detail. We give special names to several particular but important cases of
extensors. The \emph{extension,} \emph{adjoint} and \emph{generalization}
operators are introduced and their properties studied. For the so-called
$(1,1)$-extensors we define the concept of \emph{determinant}, and their
properties are investigated. Some preliminary applications of the theory of
extensors are presented in order to show the power of the new concept in
action. An useful formula for the inversion of $(1,1)$-extensors is obtained.

\end{abstract}
\tableofcontents

\section{Introduction}

In this second paper of a series of seven we introduce the so-called
\emph{extensors, }and develop some aspects of their theory with considerable
detail. We give special names and study in detail some important cases of
extensors which will appear frequently in future developments of ours
theories. The \emph{extension, adjoint} and \emph{generalization} operators
are introduced and their main properties are determined. In particular, the
generalization operator plays a crucial role in our theory of covariant
derivative operators on smooth manifolds to be presented in a forthcoming
series of papers. We define moreover the concept of determinant of a
$(1,1)$-extensor (a scalar which is an characteristic invariant associated to
the extensor), and obtain their basic properties which are very similar, but
not identical, to the classical determinat of a real square matrix. We present
some preliminary applications of the concept of extensor which will be useful
later. In particular, we present the concept of the \emph{changing basis
extensor}, and derive also an interesting formula for the inversion of
$(1,1)$-extensors. Some related, but not equivalent material, appears in (
\cite{1},\cite{3}).

\section{General $k$-Extensors}

Let $V$ be a real vector space of finite dimension, and let $V^{\ast}$ be its
dual space. Denote by $\bigwedge\nolimits^{p}V$ the space of $p$-vectors over
$V.$ Recall that if $\dim V=n$, then $\dim\bigwedge\nolimits^{p}V=\binom{n}%
{p}$.

As defined in paper I \cite{4} a multivector over $V$ is simply a formal sum
of scalar, vector,$\ldots$, pseudovector and pseudoscalar. The space of these
objects has been denoted by $\bigwedge V$, i.e., $\bigwedge V=\mathbb{R}%
+V+\cdots+\bigwedge^{n-1}V+\bigwedge^{n}V$. Recall that if $\dim V=n$ then
$\dim\bigwedge V=2^{n}$.

Let $\bigwedge\limits_{1}^{\diamond}V,\ldots,\bigwedge\limits_{k}^{\diamond}V
$ to $\bigwedge\limits^{\diamond}V$ be $k+1$ subspaces of $\bigwedge V$ such
that each of them is any sum of homogeneous subspaces of $\bigwedge V,$ and
$\bigwedge\limits^{\diamond}V$ be either any sum of homogeneous subspaces of
$\bigwedge V$ or the trivial subspace $\{0\}.$ A multilinear mapping from the
cartesian product $\bigwedge\limits_{1}^{\diamond}V\times\cdots\times
\bigwedge\limits_{k}^{\diamond}V$ to $\bigwedge\limits^{\diamond}V$ will be
called a \emph{general} $k$-\emph{extensor} over $V,$ i.e., $t:\bigwedge
\limits_{1}^{\diamond}V\times\cdots\times\bigwedge\limits_{k}^{\diamond}V$
$\rightarrow$ $\bigwedge\limits^{\diamond}V$ such that for any $\alpha
_{j},\alpha_{j}^{\prime}\in\mathbb{R}$ and $X_{j},X_{j}^{\prime}\in
\bigwedge\limits_{j}^{\diamond}V$
\begin{equation}
t(\ldots,\alpha_{j}X_{j}+\alpha_{j}^{\prime}X_{j}^{\prime},\ldots)=\alpha
_{j}t(\ldots,X_{j},\ldots)+\alpha_{j}^{\prime}t(\ldots,X_{j}^{\prime}%
,\ldots),\label{GE.1}%
\end{equation}
for each $j$ with $1\leq j\leq k.$

It should be noticed that the linear operators on $V,$ $\bigwedge^{p}V$ or
$\bigwedge V$ which appear in ordinary linear algebra are particular cases of
$1$-extensors over $V$. A covariant $k$-tensor over $V$ is just being a
$k$-extensor over $V$.

In this way, the concept of general $k$-extensor generalizes and unifies both
of the well-known concepts of linear operator and covariant $k$-tensor. These
mathematical objects are of the same nature!

The set of general $k$-extensors over $V,$ denoted by the suggestive notation
$k$-$ext(\bigwedge\limits_{1}^{\diamond}V,\cdots,\bigwedge\limits_{k}%
^{\diamond}V,\bigwedge\limits^{\diamond}V),$ has a natural structure of real
vector space. Its dimension is given by noticeable formula
\begin{equation}
\dim k\text{-}ext(\bigwedge_{1}^{\diamond}V,\ldots,\bigwedge_{k}^{\diamond
}V;\bigwedge^{\diamond}V)=\dim\bigwedge_{1}^{\diamond}V\cdots\dim\bigwedge
_{k}^{\diamond}V\dim\bigwedge^{\diamond}V.\label{GE.2}%
\end{equation}
We shall need to consider only some particular cases of these general
$k$-extensors over $V.$ So, special names and notations will be given for them.

We will equip $V$ with an arbitrary (but fixed, once and for all) euclidean
metric $G_{E}.$ And as usual we will denote the scalar product of multivectors
$X,Y\in\bigwedge V$ with respect to the euclidean metric structure
$(V,G_{E}),$ namely $X\underset{G_{E}}{\cdot}Y,$ by the more simple notation
$X\cdot Y.$

Let $\{e_{j}\}$ be any basis for $V,$ and $\{e^{j}\}$ be its euclidean
reciprocal basis for $V,$ i.e., $e_{j}\cdot e^{k}=\delta_{j}^{k}$.

\subsection{$(p,q)$-Extensors}

Let $p$ and $q$ be two integer numbers with $0\leq p,q\leq n$. A linear
mapping which sends $p$-vectors to $q$-vectors will be called a $(p,q)$%
-\emph{extensor} over $V.$ The space of them, namely $1$-$ext(\bigwedge
^{p}V;\bigwedge^{q}V),$ will be denoted by $ext_{p}^{q}(V)$ for short. By
using eq.(\ref{GE.2}) we get
\begin{equation}
\dim ext_{p}^{q}(V)=\binom{n}{p}\binom{n}{q}.\label{GE.3}%
\end{equation}

For instance, we see that the $(1,1)$-extensors over $V$ are just the linear
operators on $V.$

The set of $\binom{n}{p}\binom{n}{q}$ extensors belonging to $ext_{p}^{q}(V),
$ namely $\varepsilon^{j_{1}\ldots j_{p};k_{1}\ldots k_{q}}, $ defined by
\begin{equation}
\varepsilon^{j_{1}\ldots j_{p};k_{1}\ldots k_{q}}(X)=(e^{j_{1}}\wedge
\ldots\wedge e^{j_{p}})\cdot Xe^{k_{1}}\wedge\ldots\wedge e^{k_{q}%
},\label{GE.3a}%
\end{equation}
for all $X\in\bigwedge^{p}V,$ is a $(p,q)$-extensor basis for $ext_{p}%
^{q}(V).$

This set of extensors are indeed linearly independent, and for each $t\in
ext_{p}^{q}(V)$ there exist $\binom{n}{p}\binom{n}{q}$ real numbers, say
$t_{j_{1}\ldots j_{p};k_{1}\ldots k_{q}},$ given by
\begin{equation}
t_{j_{1}\ldots j_{p};k_{1}\ldots k_{q}}=t(e_{j_{1}}\wedge\ldots\wedge
e_{j_{p}})\cdot(e_{k_{1}}\wedge\ldots\wedge e_{k_{q}})\label{GE.3b}%
\end{equation}
such that
\begin{equation}
t=\frac{1}{p!}\frac{1}{q!}t_{j_{1}\ldots j_{p};k_{1}\ldots k_{q}}%
\varepsilon^{j_{1}\ldots j_{p};k_{1}\ldots k_{q}}.\label{GE.3c}%
\end{equation}
Such $t_{j_{1}\ldots j_{p};k_{1}\ldots k_{q}}$ will be called the $j_{1}\ldots
j_{p};k_{1}\ldots k_{q}$-th \emph{covariant components} of $t$ with respect to
the $(p,q)$-\emph{extensor basis} $\{\varepsilon^{j_{1}\ldots j_{p}%
;k_{1}\ldots k_{q}}\}.$

Of course, there are still other $(p,q)$-extensor bases for $ext_{p}^{q}(V)$
besides the ones given by eq.(\ref{GE.3a}) which can be constructed with the
vector bases $\{e_{j}\}$ and $\{e^{j}\}$ of $V$. Indeed, there are $2^{p+q}$
of such $(p,q)$-extensor bases for $ext_{p}^{q}(V).$ For instance, if we use
the basis $(p,q)$-extensors $\varepsilon_{j_{1}\ldots j_{p};k_{1}\ldots k_{q}%
}$ and the real numbers $t^{j_{1}\ldots j_{p};k_{1}\ldots k_{q}}$ defined by
\begin{align}
\varepsilon_{j_{1}\ldots j_{p};k_{1}\ldots k_{q}}(X)  &  =(e_{j_{1}}%
\wedge\ldots\wedge e_{j_{p}})\cdot Xe_{k_{1}}\wedge\ldots\wedge e_{k_{q}%
},\label{GE.3d}\\
t^{j_{1}\ldots j_{p};k_{1}\ldots k_{q}}  &  =t(e^{j_{1}}\wedge\ldots\wedge
e^{j_{p}})\cdot(e^{k_{1}}\wedge\ldots\wedge e^{k_{q}}),\label{GE.3e}%
\end{align}
we get an expansion formula for $t\in ext_{p}^{q}(V)$ analogous to that given
by eq.(\ref{GE.3c}), i.e.,
\begin{equation}
t=\frac{1}{p!}\frac{1}{q!}t^{j_{1}\ldots j_{p};k_{1}\ldots k_{q}}%
\varepsilon_{j_{1}\ldots j_{p};k_{1}\ldots k_{q}}.\label{GE.3f}%
\end{equation}
Such $t^{j_{1}\ldots j_{p};k_{1}\ldots k_{q}}$ are called the $j_{1}\ldots
j_{p};k_{1}\ldots k_{q}$-th \emph{contravariant components} of $t$ with
respect to the $(p,q)$-\emph{extensor basis} $\{\varepsilon_{j_{1}\ldots
j_{p};k_{1}\ldots k_{q}}\}.$

\subsection{Extensors}

A linear mapping which sends multivectors to multivectors will be simply
called an \emph{extensor} over $V.$ They are the linear operators on
$\bigwedge V.$ For the space of extensors over $V,$ namely $1$-$ext(\bigwedge
V;\bigwedge V),$ we will use the short notation $ext(V).$ By using
eq.(\ref{GE.2}) we get
\begin{equation}
\dim ext(V)=2^{n}2^{n}.\label{GE.4}%
\end{equation}

For instance, we will see that the so-called Hodge star operator is just being
a well-defined extensor over $V$ which can be also thought as $(p,n-p)$%
-extensor over $V.$ The so-called extended of $t\in ext_{1}^{1}(V)$ is just
being an extensor over $V,$ i.e., $\underline{t}\in ext(V).$

There are $2^{n}2^{n}$ extensors over $V,$ namely $\varepsilon^{J;K},$ given
by\footnote{$J$ and $K$ are colective indices. Recall, for example, that:
$e_{J}=1,e_{j_{1}},e_{j_{1}}\wedge e_{j_{2}},\ldots$($e^{J}=1,e^{j_{1}%
},e^{j_{1}}\wedge e^{j_{2}},\ldots$) and $\nu(J)=0,1,2,\ldots$ for
$J=\emptyset,j_{1},j_{1}j_{2},\ldots,$ where all index $j_{1},j_{2},\ldots$
runs from $1$ to $n.$}
\begin{equation}
\varepsilon^{J;K}(X)=(e^{J}\cdot X)e^{K},\label{GE.4a}%
\end{equation}
for all $X\in\bigwedge V,$ which set extensor bases for $ext(V).$

In fact they are lineary independent, and for each $t\in ext(V)$ there exist
$2^{n}2^{n}$ real numbers, say $t_{J;K},$ given by
\begin{equation}
t_{J;K}=t(e_{J})\cdot e_{K}\label{GE.4b}%
\end{equation}
such that
\begin{equation}
t=\underset{J}{\sum}\underset{K}{\sum}\frac{1}{\nu(J)!}\frac{1}{\nu
(K)!}t_{J;K}\varepsilon^{J;K}.\label{GE.4c}%
\end{equation}
Such $t_{J;K}$ will be called the $J;K$-th\emph{\ covariant components} of $t
$ with respect to the \emph{extensor basis} $\{\varepsilon^{J;K}\}.$

We notice that exactly $(2^{n+1}-1)^{2}$ extensor bases for $ext(V)$ can be
constructed from the vector bases $\{e_{j}\}$ and $\{e^{j}\}$ of $V.$ For
instance, whenever the basis extensors $\varepsilon_{J;K}$ and the real
numbers $t^{J;K}$ defined by
\begin{align}
\varepsilon_{J;K}(X)  &  =(e_{J}\cdot X)e_{K},\label{GE.4d}\\
t^{J;K}  &  =t(e^{J})\cdot e^{K}\label{GE.4e}%
\end{align}
are being used, an expansion formula for $t\in ext(V)$ analogous to that given
by eq.(\ref{GE.4c}) can be obtained, i.e.,
\begin{equation}
t=\underset{J}{\sum}\underset{K}{\sum}\frac{1}{\nu(J)!}\frac{1}{\nu
(K)!}t^{J;K}\varepsilon_{J;K}.\label{GE.4f}%
\end{equation}
Such $t^{J;K}$ are called the $J;K$-th \emph{contravariant components} of $t$
with respect to the \emph{extensor basis} $\{\varepsilon_{J;K}\}$.

\subsection{Elementary $k$-Extensors}

A multilinear mapping which takes $k$-uple of vectors into $q$-vectors will be
called an \emph{elementary }$k$-\emph{extensor} over $V$of degree $q.$ The
space of them, namely $k$-$ext(V,\ldots,V;\bigwedge^{q}V),$ will be denoted by
$k$-$ext^{q}(V).$ It is easy to verify by using eq.(\ref{GE.2}) that
\begin{equation}
\dim k\text{-}ext^{q}(V)=n^{k}\binom{n}{q}.\label{GE.5}%
\end{equation}
It should be noticed that a\emph{\ }elementary $k$-extensor over $V$ of degree
$0$ is just being a \emph{covariant }$k$-\emph{tensor} over $V,$ i.e.,
$k$-$ext^{0}(V)\equiv T_{k}(V).$ It is easily realized that $1$-$ext^{q}%
(V)\equiv ext_{1}^{q}(V).$

The elementary $k$-extensors of degrees $0,1,2,\ldots$ etc. are sometimes said
to be \emph{scalar}, \emph{vector,} \emph{bivector}, $\ldots$ etc.
\emph{elementary }$k$-\emph{extensors.}

There are $n^{k}\binom{n}{q}$ elementary $k$-extensors of degree $q$ belonging
to $k$-$ext^{q}(V),$ namely $\varepsilon^{j_{1},\ldots,j_{k};k_{1}\ldots
k_{q}},$ defined by
\begin{equation}
\varepsilon^{j_{1},\ldots,j_{k};k_{1}\ldots k_{q}}(v_{1},\ldots,v_{k}%
)=(v_{1}\cdot e^{j_{1}})\ldots(v_{k}\cdot e^{j_{k}})e^{k_{1}}\wedge
\ldots\wedge e^{k_{q}},\label{GE.5a}%
\end{equation}
for all $(v_{1},\ldots,v_{k})\in\underset{k\text{ copies}}{\underbrace
{V\times\cdots\times V}},$ which set elementary $k$-extensor of degree $q$
bases for $k$-$ext^{q}(V).$

In fact they are linearly independent, and for all $t\in k$-$ext^{q}(V)$ there
are $n^{k}\binom{n}{q}$ real numbers, say $t_{j_{1},\ldots,j_{k};k_{1}\ldots
k_{q}},$ given by
\begin{equation}
t_{j_{1},\ldots,j_{k};k_{1}\ldots k_{q}}=t(e_{j_{1}},\ldots,e_{j_{k}}%
)\cdot(e_{k_{1}}\wedge\ldots\wedge e_{k_{q}})\label{GE.5b}%
\end{equation}
such that
\begin{equation}
t=\frac{1}{q!}t_{j_{1},\ldots,j_{k};k_{1}\ldots k_{q}}\varepsilon
^{j_{1},\ldots,j_{k};k_{1}\ldots k_{q}}.\label{GE.5c}%
\end{equation}
Such $t_{j_{1},\ldots,j_{k};k_{1}\ldots k_{q}}$ will be called the
$j_{1},\ldots,j_{k};k_{1}\ldots k_{q}$-th \emph{covariant components }of $t$
with respect to the\emph{\ bases} $\{\varepsilon^{j_{1},\ldots,j_{k}%
;k_{1}\ldots k_{q}}\}.$

We notice that exactly $2^{k+q}$ elementary $k$-extensors of degree $q$ bases
for $k$-$ext^{q}(V)$ can be constructed from the vector bases $\{e_{j}\}$ and
$\{e^{j}\}$ of $V$. For instance, we might define the basis elementary
$k$-extensor of degree $q$ $\varepsilon_{j_{1},\ldots,j_{k};k_{1}\ldots k_{q}%
}$ and the real numbers $t^{j_{1},\ldots,j_{k};k_{1}\ldots k_{q}} $ by the
following equations
\begin{align}
\varepsilon_{j_{1},\ldots,j_{k};k_{1}\ldots k_{q}}(v_{1},\ldots,v_{k})  &
=(v_{1}\cdot e_{j_{1}})\ldots(v_{k}\cdot e_{j_{k}})e_{k_{1}}\wedge\ldots\wedge
e_{k_{q}},\label{GE.5d}\\
t^{j_{1},\ldots,j_{k};k_{1}\ldots k_{q}}  &  =t(e^{j_{1}},\ldots,e^{j_{k}%
})\cdot(e^{k_{1}}\wedge\ldots\wedge e^{k_{q}}).\label{GE.5e}%
\end{align}
Then, we might have other expansion formula for $t\in k$-$ext^{q}(V)$ than
that given by eq.(\ref{GE.5c}), i.e.,
\begin{equation}
t=\frac{1}{q!}t^{j_{1},\ldots,j_{k};k_{1}\ldots k_{q}}\varepsilon
_{j_{1},\ldots,j_{k};k_{1}\ldots k_{q}}.\label{GE.5f}%
\end{equation}
Such $t^{j_{1},\ldots,j_{k};k_{1}\ldots k_{q}}$ are called the $j_{1}%
,\ldots,j_{k};k_{1}\ldots k_{q}$-th \emph{contravariant components} of $t$
with respect to the \emph{basis} $\{\varepsilon_{j_{1},\ldots,j_{k}%
;k_{1}\ldots k_{q}}\}.$

Note that $1$-$ext^{p}(V)\equiv ext_{1}^{p}(V)$.

A completely skew-symmetric $k$-extensor over $V$ of degree $p$ ($k\geq2$),
i.e., $\Theta\in k$-$ext^{p}(V)$ such that for any $v_{i},v_{j}\in V$ with
$1\leq i<j\leq k,$
\begin{equation}
\Theta(v_{1},\ldots,v_{i},\ldots,v_{j},\ldots,v_{k})=-\Theta(v_{1}%
,\ldots,v_{j},\ldots,v_{i},\ldots,v_{k})\label{2.28b}%
\end{equation}
will be called a $k$\emph{-exform over }$V$\emph{\ of degree }$p.$

The vector space of $k$-exforms over $V$ of degree $p$ will be denoted by
$k$-$exf^{p}(V).$

It is also convenient to define a $0$-exform of degree $p$ to be a $p$-vector
(i.e., $0$-$exf^{p}(V)=\bigwedge^{p}V$) and an $1$-exform of degree $p$ to be
a $1$-extensor of degree $p$ (i.e., $1$-$exf^{p}(V)=1$-$ext^{p}(V)$).

If $\dim V=n,$ then $\dim k$-$exf^{p}(V)=\binom{n}{k}\binom{n}{p}.$

The $k$-exforms of degree $0,1,2,\ldots$ etc. are said to be `scalar
$k$-exforms', vector $k$-exforms', `bivector $k$-exforms',$\ldots$\ etc.

Note that a scalar $k$-exform is just a $k$-form, i.e., $k$-$exf^{0}%
(V)=\bigwedge^{k}V.$ In this way, we see that the concept of $k$-exform
generalizes the concept of $k$-form.

\section{Projectors}

Let $\bigwedge\limits^{\diamond}V$ be either any sum of homogeneous
subspaces\footnote{Note that for such a subspace $\bigwedge\limits^{\diamond
}V$ there are $\nu$ integers $p_{1,}\ldots,p_{\nu}$ ($0\leq p_{1}%
<\cdots<p_{\nu}\leq n$) such that $\bigwedge\limits^{\diamond}V=\bigwedge
^{p_{1}}V+\cdots+\bigwedge^{p_{\nu}}V.$} of $\bigwedge V$ or the trivial
subspace $\{0\}.$ Associated to $\bigwedge^{\diamond}V,$ a noticeable extensor
from $\bigwedge V$ to $\bigwedge\limits^{\diamond}V,$ namely $\left\langle
\left.  {}\right.  \right\rangle _{\bigwedge\limits^{\diamond}V},$ can defined
by
\begin{equation}
\left\langle X\right\rangle _{\bigwedge\limits^{\diamond}V}=\left\{
\begin{array}
[c]{cc}%
\left\langle X\right\rangle _{p_{1}}+\cdots+\left\langle X\right\rangle
_{p_{\nu}}, & \text{if }\bigwedge\limits^{\diamond}V=\bigwedge^{p_{1}}%
V+\cdots+\bigwedge^{p_{\nu}}V\\
0, & \text{if }\bigwedge\limits^{\diamond}V=\{0\}
\end{array}
\right.  .\label{P.1}%
\end{equation}
Such $\left\langle \left.  {}\right.  \right\rangle _{\bigwedge
\limits^{\diamond}V}\in1$-$ext(\bigwedge V;\bigwedge\limits^{\diamond}V)$ will
be called the $\bigwedge^{\diamond}V$-\emph{projector extensor}.

We notice that if $\bigwedge\limits^{\diamond}V$ is any homogeneous subspace
of $\bigwedge V,$ i.e., $\bigwedge\limits^{\diamond}V=\bigwedge^{p}V,$ then
the projector extensor is reduced to the so-called $p$-\emph{part operator},
i.e., $\left\langle \left.  {}\right.  \right\rangle _{\bigwedge
\limits^{\diamond}V}=\left\langle \left.  {}\right.  \right\rangle _{p}.$

We now summarize the fundamental properties for the $\bigwedge
\limits^{\diamond}V$-projector extensors.

Let $\bigwedge\limits_{1}^{\diamond}V$ and $\bigwedge\limits_{2}^{\diamond}V$
be two subspaces of $\bigwedge V.$ If each of them is either any sum of
homogeneous subspaces of $\bigwedge V$ or the trivial subspace $\{0\},$ then
\begin{align}
\left\langle \left\langle X\right\rangle _{\bigwedge\limits_{1}^{\diamond}%
V}\right\rangle _{\bigwedge\limits_{2}^{\diamond}V}  &  =\left\langle
X\right\rangle _{\bigwedge\limits_{1}^{\diamond}V\cap\bigwedge\limits_{2}%
^{\diamond}V}\label{P.2a}\\
\left\langle X\right\rangle _{\bigwedge\limits_{1}^{\diamond}V}+\left\langle
X\right\rangle _{\bigwedge\limits_{2}^{\diamond}V}  &  =\left\langle
X\right\rangle _{\bigwedge\limits_{1}^{\diamond}V\cup\bigwedge\limits_{2}%
^{\diamond}V}.\label{P.2b}%
\end{align}

Let $\bigwedge\limits^{\diamond}V$ be either any sum of homogeneous subspaces
of $\bigwedge V$ or the trivial subspace $\{0\}.$ It holds
\begin{equation}
\left\langle X\right\rangle _{\bigwedge\limits^{\diamond}V}\cdot
Y=X\cdot\left\langle Y\right\rangle _{\bigwedge\limits^{\diamond}%
V}.\label{P.2c}%
\end{equation}

In this sense we might say that the concept of $\bigwedge\limits^{\diamond}V
$-projector extensor is just a well-done generalization of the concept of
$p$-part operator.

\section{The Extension Operator}

Let $\{e_{j}\}$ be any basis for $V,$ and $\{\varepsilon^{j}\}$ be its dual
basis for $V^{*}.$ As we know, $\{\varepsilon^{j}\}$ is the unique $1$-form
basis associated to the vector basis $\{e_{j}\}$ such that $\varepsilon
^{j}(e_{i})=\delta_{i}^{j}.$

The mapping\footnote{Observe that the extension operator is \textit{not} a
linear mapping.} $ext_{1}^{1}(V)\ni t\mapsto\underline{t}\in ext(V)$ such that
for any $X\in\bigwedge V:$ if $X=X_{0}+\overset{n}{\underset{k=1}{%
{\displaystyle\sum}
}}X_{k},$ then
\begin{equation}
\underline{t}(X)=X_{0}+\overset{n}{\underset{k=1}{%
{\displaystyle\sum}
}}\frac{1}{k!}X_{k}(\varepsilon^{j_{1}},\ldots,\varepsilon^{j_{k}})t(e_{j_{1}%
})\wedge\ldots\wedge t(e_{j_{k}})\label{EO.1}%
\end{equation}
will be called the \emph{extension operator}. We call $\underline{t}$ the
\emph{extended} of $t.$ It is the well-known \emph{outermorphism} of $t$ in
the ordinary linear algebra.

The extension operator is well-defined since it does not depend on the choice
of $\{e_{j}\}.$

We summarize now the basic properties satisfied by the extension operator.

\textbf{e1} The extension operator is grade-preserving, i.e.,
\begin{equation}
\text{if }X\in\bigwedge\nolimits^{p}V,\text{ then }\underline{t}%
(X)\in\bigwedge\nolimits^{p}V.\label{EO.1a}%
\end{equation}
It is an obvious result which follows from eq.(\ref{EO.1}).

\textbf{e2} For any $\alpha\in\mathbb{R},$ $v\in V$ and $v_{1}\wedge
\ldots\wedge v_{k}\in\bigwedge^{k}V$ it holds
\begin{align}
\underline{t}(\alpha)  &  =\alpha,\label{EO.2a}\\
\underline{t}(v)  &  =t(v),\label{EO.2b}\\
\underline{t}(v_{1}\wedge\ldots\wedge v_{k})  &  =t(v_{1})\wedge\ldots\wedge
t(v_{k}).\label{EO.2c}%
\end{align}

\begin{proof}
The first statement trivially follows from eq.(\ref{EO.1}). The second one can
easily be deduced from eq.(\ref{EO.1}) by recalling the elementary expansion
formula for vectors and the linearity of extensors. In order to prove the
third statement we will use the remarkable formulas: $v_{1}\wedge\ldots\wedge
v_{k}(\omega^{1},\ldots,\omega^{k})=\epsilon^{i_{1}\ldots i_{k}}\omega
^{1}(v_{i_{1}})\ldots\omega^{k}(v_{i_{k}})$ and $w_{i_{1}}\wedge\ldots\wedge
w_{i_{k}}=\epsilon_{i_{1}\ldots i_{k}}w_{1}\wedge\ldots\wedge w_{k},$ where
$v_{1},\ldots,v_{k}\in V,$ $w_{1},\ldots,w_{k}\in V$ and $\omega^{1}%
,\ldots,\omega^{k}\in V^{*},$ and the combinatorial formula $\epsilon
^{i_{1}\ldots i_{k}}\epsilon_{i_{1}\ldots i_{k}}=k!.$ By recalling the
elementary expansion formula for vectors and the linearity of extensors we
have that
\begin{align*}
\underline{t}(v_{1}\wedge\ldots\wedge v_{k})  &  =\frac{1}{k!}v_{1}%
\wedge\ldots\wedge v_{k}(\varepsilon^{j_{1}},\ldots,\varepsilon^{j_{k}%
})t(e_{j_{1}})\wedge\ldots\wedge t(e_{j_{k}})\\
&  =\frac{1}{k!}\epsilon^{i_{1}\ldots i_{k}}\varepsilon^{j_{1}}(v_{i_{1}%
})\ldots\varepsilon^{j_{k}}(v_{i_{k}})t(e_{j_{1}})\wedge\ldots\wedge
t(e_{j_{k}})\\
&  =\frac{1}{k!}\epsilon^{i_{1}\ldots i_{k}}t(v_{i_{1}})\wedge\ldots\wedge
t(v_{i_{k}}),\\
&  =t(v_{1})\wedge\ldots\wedge t(v_{k}).
\end{align*}
\end{proof}

\textbf{e3} For any $X,Y\in\bigwedge V$ it holds
\begin{equation}
\underline{t}(X\wedge Y)=\underline{t}(X)\wedge\underline{t}(Y).\label{EO.3}%
\end{equation}
It is an immediate result which follows from eq.(\ref{EO.2c}).

We emphasize that the three fundamental properties as given by eq.(\ref{EO.2a}%
), eq.(\ref{EO.2b}) and eq.(\ref{EO.3}) together are completely equivalent to
the \emph{extension procedure} as defined by eq.(\ref{EO.1}).

We present next some important properties of the extension operator.

\textbf{e4} Let us take $s,t\in ext_{1}^{1}(V).$ Then, the following result
holds
\begin{equation}
\underline{s\circ t}=\underline{s}\circ\underline{t}.\label{EO.4}%
\end{equation}

\begin{proof}
It is enough to present the proofs for scalars and simple $k$-vectors.
For $\alpha\in\mathbb{R},$ by using eq.(\ref{EO.2a}), we get
\[
\underline{s\circ t}(\alpha)=\alpha=\underline{s}(\alpha)=\underline
{s}(\underline{t}(\alpha))=\underline{s}\circ\underline{t}(\alpha).
\]
For a simple $k$-vector $v_{1}\wedge\ldots\wedge v_{k}\in\bigwedge^{k}V,$ by
using eq.(\ref{EO.2c}), we get
\begin{align*}
\underline{s\circ t}(v_{1}\wedge\ldots\wedge v_{k})  &  =s\circ t(v_{1}%
)\wedge\ldots\wedge s\circ t(v_{k})=s(t(v_{1}))\wedge\ldots\wedge
s(t(v_{k}))\\
&  =\underline{s}(t(v_{1})\wedge\ldots\wedge t(v_{k}))=\underline
{s}(\underline{t}(v_{1}\wedge\ldots\wedge v_{k})),\\
&  =\underline{s}\circ\underline{t}(v_{1}\wedge\ldots\wedge v_{k}).
\end{align*}
Next we can easily generalize to multivectors by linearity of extensors. It
yields
\[
\underline{s\circ t}(X)=\underline{s}\circ\underline{t}(X).
\]
\end{proof}

\textbf{e5} Let us take $t\in ext_{1}^{1}(V)$ with inverse $t^{-1}\in
ext_{1}^{1}(V),$ i.e., $t^{-1}\circ t=t\circ t^{-1}=i_{V}.$ Then,
$\underline{(t^{-1})}\in ext(V)$ is the inverse of $\underline{t}\in ext(V),$
i.e.,
\begin{equation}
(\underline{t})^{-1}=\underline{(t^{-1})}.\label{EO.5}%
\end{equation}
Indeed, by using eq.(\ref{EO.4}) and the obvious property $\underline{i}%
_{V}=i_{\bigwedge V},$ we have that
\[
t^{-1}\circ t=t\circ t^{-1}=i_{V}\Rightarrow\underline{(t^{-1})}%
\circ\underline{t}=\underline{t}\circ\underline{(t^{-1})}=i_{\bigwedge V}.
\]
It means that the \emph{inverse of the extended} of $t$ equals the
\emph{extended of the inverse} of $t.$

In accordance with the corollary above we might use a more simple notation
$\underline{t}^{-1}$ to denote both of $(\underline{t})^{-1}$ and
$\underline{(t^{-1})}.$

\textbf{e6} Let $\{e_{j}\}$ be any basis for $V,$ and $\{e^{j}\}$ its
euclidean reciprocal basis for $V,$ i.e., $e_{j}\cdot e^{k}=\delta_{j}^{k}.$
There are two interesting and useful formulas for calculating the extended of
$t\in ext_{1}^{1}(V),$ i.e.,%

\begin{align}
\underline{t}(X)  &  =1\cdot X+\overset{n}{\underset{k=1}{\sum}}\frac{1}%
{k!}(e^{j_{1}}\wedge\ldots\wedge e^{j_{k}})\cdot Xt(e_{j_{1}})\wedge
\ldots\wedge t(e_{j_{k}})\label{EO.6a}\\
&  =1\cdot X+\overset{n}{\underset{k=1}{\sum}}\frac{1}{k!}(e_{j_{1}}%
\wedge\ldots\wedge e_{j_{k}})\cdot Xt(e^{j_{1}})\wedge\ldots\wedge t(e^{j_{k}%
}).\label{EO.6b}%
\end{align}

\section{Standard Adjoint Operator}

Let as above $\bigwedge\limits_{1}^{\diamond}V$ and $\bigwedge\limits_{2}%
^{\diamond}V$ be two subspaces of $\bigwedge V$ such that each of them is
either any sum of homogeneous subspaces of $\bigwedge V.$ Let $\{e_{j}\}$ and
$\{e^{j}\}$ be two euclidean reciprocal bases to each other for $V,$ i.e.,
$e_{j}\cdot e^{k}=\delta_{j}^{k}.$

The \emph{standard adjoint operator} is the linear mapping $1$-$ext(\bigwedge
\limits_{1}^{\diamond}V;\bigwedge\limits_{2}^{\diamond}V)\ni t\rightarrow
t^{\dagger}\in1$-$ext(\bigwedge\limits_{2}^{\diamond}V;\bigwedge
\limits_{1}^{\diamond}V)$ such that for any $Y\in\bigwedge\limits_{2}%
^{\diamond}V:$%
\begin{align}
t^{\dagger}(Y)  &  =t(\left\langle 1\right\rangle _{\bigwedge\limits_{1}%
^{\diamond}V})\cdot Y+\underset{k=1}{\overset{n}{\sum}}\frac{1}{k!}%
t(\left\langle e^{j_{1}}\wedge\ldots e^{j_{k}}\right\rangle _{\bigwedge
\limits_{1}^{\diamond}V})\cdot Ye_{j_{1}}\wedge\ldots e_{j_{k}}\label{SAO.1a}%
\\
&  =t(\left\langle 1\right\rangle _{\bigwedge\limits_{1}^{\diamond}V})\cdot
Y+\underset{k=1}{\overset{n}{\sum}}\frac{1}{k!}t(\left\langle e_{j_{1}}%
\wedge\ldots e_{j_{k}}\right\rangle _{\bigwedge\limits_{1}^{\diamond}V})\cdot
Ye^{j_{1}}\wedge\ldots e^{j_{k}},\label{SAO.1b}%
\end{align}
or into a more compact notation by employing the \emph{colective index} $J,$
\begin{align}
t^{\dagger}(Y)  &  =\underset{J}{\sum}\frac{1}{\nu(J)!}t(\left\langle
e^{J}\right\rangle _{\bigwedge_{1}^{\diamond}V})\cdot Ye_{J}\label{SAO.2a}\\
&  =\underset{J}{\sum}\frac{1}{\nu(J)!}t(\left\langle e_{J}\right\rangle
_{\bigwedge_{1}^{\diamond}V})\cdot Ye^{J},\label{SAO.2b}%
\end{align}
We call $t^{\dagger}$ the \emph{standard adjoint extensor} of $t.$ It should
be noticed the use of the $\bigwedge\limits_{1}^{\diamond}V$-projector extensor.

The standard adjoint operator is well-defined since the sums appearing in each
of places above do not depend on the choice of $\{e_{j}\}.$

Let us take $X\in\bigwedge\limits_{1}^{\diamond}V$ and $Y\in\bigwedge
\limits_{2}^{\diamond}V$. A straightforward calculation yields
\begin{align*}
X\cdot t^{\dagger}(Y)  &  =\underset{J}{\sum}\frac{1}{\nu(J)!}t(\left\langle
e^{J}\right\rangle _{\bigwedge_{1}^{\diamond}V})\cdot Y(X\cdot e_{J})\\
&  =t(\underset{J}{\sum}\frac{1}{\nu(J)!}\left\langle (X\cdot e_{J}%
)e^{J}\right\rangle _{\bigwedge_{1}^{\diamond}V})\cdot Y\\
&  =t(\left\langle X\right\rangle _{\bigwedge_{1}^{\diamond}V})\cdot Y,
\end{align*}
i.e.,
\begin{equation}
X\cdot t^{\dagger}(Y)=t(X)\cdot Y.\label{SAO.3}%
\end{equation}
It is a generalization of the well-known property which holds for linear operators.

Let us take $t\in1$-$ext(\bigwedge_{1}^{\diamond}V;\bigwedge_{2}^{\diamond}V)
$ and $u\in1$-$ext(\bigwedge_{2}^{\diamond}V;\bigwedge_{3}^{\diamond}V).$ We
can note that $u\circ t\in1$-$ext(\bigwedge_{1}^{\diamond}V;\bigwedge
_{3}^{\diamond}V)$ and $t^{\dagger}\circ u^{\dagger}\in1$-$ext(\bigwedge
_{3}^{\diamond}V;\bigwedge_{1}^{\diamond}V).$ Then, let us take $X\in
\bigwedge_{1}^{\diamond}V$ and $Z\in\bigwedge_{3}^{\diamond}V,$ by using
eq.(\ref{SAO.3}) we have that
\[
X\cdot(u\circ t)^{\dagger}(Z)=(u\circ t)(X)\cdot Z=t(X)\cdot u^{\dagger
}(Z)=X\cdot(t^{\dagger}\circ u^{\dagger})(Z).
\]
Hence, we get
\begin{equation}
(u\circ t)^{\dagger}=t^{\dagger}\circ u^{\dagger}.\label{SAO.4}%
\end{equation}

Let us take $t\in1$-$ext(\bigwedge^{\diamond}V;\bigwedge^{\diamond}V)$ with
inverse $t^{-1}\in1$-$ext(\bigwedge^{\diamond}V;\bigwedge^{\diamond}V),$ i.e.,
$t^{-1}\circ t=t\circ t^{-1}=i_{\bigwedge^{\diamond}V},$ where $i_{\bigwedge
^{\diamond}V}\in1$-$ext(\bigwedge^{\diamond}V;\bigwedge^{\diamond}V)$ is the
so-called identity function for $\bigwedge^{\diamond}V.$ By using
eq.(\ref{SAO.4}) and the obvious property $i_{\bigwedge^{\diamond}%
V}=i_{\bigwedge^{\diamond}V}^{\dagger},$ we have that
\[
t^{-1}\circ t=t\circ t^{-1}=i_{\bigwedge^{\diamond}V}\Rightarrow t^{\dagger
}\circ(t^{-1})^{\dagger}=(t^{-1})^{\dagger}\circ t^{\dagger}=i_{\bigwedge
^{\diamond}V},
\]
hence,
\begin{equation}
(t^{\dagger})^{-1}=(t^{-1})^{\dagger},\label{SAO.5}%
\end{equation}
i.e., the inverse of the adjoint of $t$ equals the adjoint of the inverse of
$t$. In accordance with the above corollary it is possible to use a more
simple symbol, say $t^{\ast},$ to denote both of $(t^{\dagger})^{-1}$ and
$(t^{-1})^{\dagger}.$

Let us take $t\in ext_{1}^{1}(V).$ We note that $\underline{t}\in ext(V)$ and
$\underline{(t^{\dagger})}\in ext(V).$ A straightforward calculation by using
eqs.(\ref{EO.6a}) and (\ref{EO.6b}) yields
\begin{align*}
\underline{(t^{\dagger})}(Y)  &  =1\cdot Y+\overset{n}{\underset{k=1}{\sum}%
}\frac{1}{k!}(e^{j_{1}}\wedge\ldots e^{j_{k}})\cdot Yt^{\dagger}(e_{j_{1}%
})\wedge\ldots t^{\dagger}(e_{j_{k}})\\
&  =1\cdot Y+\\
&  \overset{n}{\underset{k=1}{\sum}}\frac{1}{k!}(e^{j_{1}}\wedge\ldots
e^{j_{k}})\cdot Yt^{\dagger}(e_{j_{1}})\cdot e_{p_{1}}e^{p_{1}}\wedge\ldots
t^{\dagger}(e_{j_{k}})\cdot e_{p_{k}}e^{p_{k}}\\
&  =1\cdot Y+\overset{n}{\underset{k=1}{\sum}}\frac{1}{k!}(e_{j_{1}}\cdot
t(e_{p_{1}})e^{j_{1}}\wedge\ldots e_{j_{k}}\cdot t(e_{p_{k}})e^{j_{k}})\cdot
Ye^{p_{1}}\wedge\ldots e^{p_{k}}\\
&  =\underline{t}(1)\cdot Y+\overset{n}{\underset{k=1}{\sum}}\frac{1}%
{k!}\underline{t}(e_{p_{1}}\wedge\ldots e_{p_{k}})\cdot Ye^{p_{1}}\wedge\ldots
e^{p_{k}}\\
&  =(\underline{t})^{\dagger}(Y).
\end{align*}
Hence, we get
\begin{equation}
\underline{(t^{\dagger})}=(\underline{t})^{\dagger}.\label{SAO.6}%
\end{equation}
This means that the extension operator commutes with the adjoint operator. In
accordance with the property above we may use a more simple notation
$\underline{t}^{\dagger}$ to denote without ambiguities both of $\underline
{(t^{\dagger})}$ and $(\underline{t})^{\dagger}$.

In many applications the adjoint operator is used for the cases where
$\bigwedge\limits_{1}^{\diamond}V=\bigwedge\nolimits^{p}V$ and $\bigwedge
\limits_{2}^{\diamond}V=\bigwedge\nolimits^{q}V$ are homogenous subspaces of
$\bigwedge V$. In this particular case the adjoint operator is simply a linear
mapping acting on these vector spaces of extensors, namely $ext_{q}%
^{p}(V)\backepsilon t\mapsto t^{\dagger}\in ext_{p}^{q}(V).$ We have from
eq.(\ref{SAO.2a}) and eq.(\ref{SAO.2b}) the simple formulas
\begin{align}
t^{\dagger}(Y)  &  =\frac{1}{p!}\underline{t}(e_{j_{1}}\wedge\ldots e_{j_{p}%
})\cdot Y(e^{j_{1}}\wedge\ldots e^{j_{p}})\label{SAO.7a}\\
&  =\frac{1}{p!}\underline{t}(e^{_{j_{1}}}\wedge\ldots e^{_{j_{p}}})\cdot
Y(e_{j_{1}}\wedge\ldots e_{j_{p}}),\label{SAO.7b}%
\end{align}
for all $Y\in\bigwedge^{q}V,$ where Einstein's convention has been used.

\section{The Generalization Operator}

Let $\{e_{k}\}$ be any basis for $V,$ and $\{e^{k}\}$ be its euclidean
reciprocal basis for $V,$ as we know, $e_{k}\cdot e^{l}=\delta_{k}^{l}.$

The linear mapping $ext_{1}^{1}(V)\ni t\mapsto\underset{\thicksim}{t}\in
ext(V)$ such that for any $X\in\bigwedge V$
\begin{equation}
\underset{\thicksim}{t}(X)=t(e^{k})\wedge(e_{k}\lrcorner X)=t(e_{k}%
)\wedge(e^{k}\lrcorner X)\label{SGO.1}%
\end{equation}
will be called the \emph{generalization operator.} We call $\underset
{\thicksim}{t}$ the \emph{generalized of }$t.$

The generalization operator is well-defined since it does not depend on the
choice of $\{e_{k}\}.$

We present now some important properties which are satisfied by the
generalization operator.

\textbf{g1 }The generalization operator is grade-preserving, i.e.,
\begin{equation}
\text{if }X\in\bigwedge\nolimits^{k}V,\text{ then }\underset{\thicksim}%
{t}(X)\in\bigwedge\nolimits^{k}V.\label{SGO.1a}%
\end{equation}

\textbf{g2 }The grade involution $\widehat{\left.  {}\right.  }\in ext(V),$
reversion $\widetilde{\left.  {}\right.  }\in ext(V),$ and conjugation
$\overline{\left.  {}\right.  }\in ext(V)$ commute with the generalization
operator, i.e.,
\begin{align}
\underset{\thicksim}{t}(\widehat{X})  &  =\widehat{\underset{\thicksim}{t}%
(X)},\label{SGO.2a}\\
\underset{\thicksim}{t}(\widetilde{X})  &  =\widetilde{\underset{\thicksim}%
{t}(X)},\label{SGO.2b}\\
\underset{\thicksim}{t}(\overline{X})  &  =\overline{\underset{\thicksim}%
{t}(X)}.\label{SGO.2c}%
\end{align}
They are immediate consequences of the grade-preserving property.

\textbf{g3 }For any $\alpha\in\mathbb{R},$ $v\in V$ and $X,Y\in\bigwedge V $
it holds
\begin{align}
\underset{\thicksim}{t}(\alpha)  &  =0,\label{SGO.3a}\\
\underset{\thicksim}{t}(v)  &  =t(v),\label{SGO.3b}\\
\underset{\thicksim}{t}(X\wedge Y)  &  =\underset{\thicksim}{t}(X)\wedge
Y+X\wedge\underset{\thicksim}{t}(Y).\label{SGO.3c}%
\end{align}

The proof of eq.(\ref{SGO.3a}) and eq.(\ref{SGO.3b}) are left to the reader.
Hint: $v\lrcorner\alpha=0$ and $v\lrcorner w=v\cdot w.$ Now, the identities:
$a\lrcorner(X\wedge Y)=(a\lrcorner X)\wedge Y+\widehat{X}\wedge(a\lrcorner Y)
$ and $a\wedge X=\widehat{X}\wedge a,$ with $a\in V$ and $X,Y\in\bigwedge V,$
allow us to prove the property given by eq.(\ref{SGO.3c}).

We can prove that the basic properties given by eq.(\ref{SGO.3a}),
eq.(\ref{SGO.3b}) and eq.(\ref{SGO.3c}) together are completely equivalent to
the \emph{generalization procedure} as defined by eq.(\ref{SGO.1}).

\textbf{g4} The generalization operator commutes with the adjoint operator,
i.e.,
\begin{equation}
(\underset{\thicksim}{t})^{\dagger}=\underset{\thicksim}{(t^{\dagger}%
)},\label{SGO.4}%
\end{equation}
or put it on another way, the \emph{adjoint of the generalized} of $t$ is just
the \emph{generalized of the adjoint} of $t.$

\begin{proof}
A straightforward calculation, by using eq.(\ref{SAO.3})
and the multivector identities: $X\cdot(a\wedge Y)=(a\lrcorner
X)\wedge Y$ and $X\cdot(a\lrcorner Y)=(a\wedge X)\cdot Y,$ with
$a\in V$ and $X,Y\in\bigwedge V,$ gives
\begin{align*}
(\underset{\thicksim}{t})^{\dagger}(X)\cdot Y  &  =X\cdot\underset{\thicksim
}{t}(Y)\\
&  =(e_{j}\wedge(t(e^{j})\lrcorner X))\cdot Y=(e_{j}\wedge(t(e^{j})\cdot
e^{k}e_{k}\lrcorner X))\cdot Y\\
&  =(e^{j}\cdot t^{\dagger}(e^{k})e_{j}\wedge(e_{k}\lrcorner X))\cdot
Y=(t^{\dagger}(e^{k})\wedge(e_{k}\lrcorner X))\cdot Y\\
&  =\underset{\thicksim}{(t^{\dagger})}(X)\cdot Y.
\end{align*}
Hence, by non-degeneracy of the euclidean scalar product, the required result
follows.
\end{proof}
In accordance with the above property we might use a more simple symbol
$\underset{\thicksim}{t^{\dagger}}$ to mean $(\underset{\thicksim}%
{t})^{\dagger}$ and $\underset{\thicksim}{(t^{\dagger})}.$

\textbf{g5} The symmetric (skew-symmetric) part of the generalized of $t$ is
just the generalized of the symmetric (skew-symmetric) part of $t,$ i.e.,
\begin{equation}
(\underset{\thicksim}{t})_{\pm}=(\underset{\thicksim}{t_{\pm}}).\label{SGO.5}%
\end{equation}
This property follows immediately from eq.(\ref{SGO.4}).

We see also that it is possible to use the more simple notation $\underset
{\thicksim}{t}_{\pm}$ to mean $(\underset{\thicksim}{t})_{\pm}$ and
$(\underset{\thicksim}{t_{\pm}}).$

\textbf{g6} The skew-symmetric part of the generalized of $t$ can be
factorized by the noticeable formula\footnote{Recall that $X\times
Y\equiv\frac{1}{2}(XY-YX).$}
\begin{equation}
\underset{\thicksim}{t}_{-}(X)=\frac{1}{2}biv[t]\times X,\label{SGO.6}%
\end{equation}
where $biv[t]\equiv t(e^{k})\wedge e_{k}$ is an \emph{characteristic
invariant} of $t,$ the so-called \emph{bivector of} $t.$

\begin{proof}
By using eq.(\ref{SGO.5}), the well-known identity $t_{-}(a)=\frac{1}%
{2}biv[t]\times a$ and the remarkable multivector identity $B\times X=(B\times
e^{k})\wedge(e_{k}\lrcorner X),$ with $B\in\bigwedge^{2}V$ and $X\in\bigwedge
V,$ we have that
\[
\underset{\thicksim}{t}_{-}(X)=t_{-}(e^{k})\wedge(e_{k}\lrcorner X)=(\frac
{1}{2}biv[t]\times e^{k})\wedge(e_{k}\lrcorner X)=\frac{1}{2}biv[t]\times X.
\]
\end{proof}
\textbf{g7} A noticeable formula holds for the skew-symmetric part of the
generalized of $t.$ For all $X,Y\in\bigwedge V$
\begin{equation}
\underset{\thicksim}{t}_{-}(X*Y)=\underset{\thicksim}{t}_{-}(X)*Y+X*\underset
{\thicksim}{t}_{-}(Y),\label{SGO.7}%
\end{equation}
where $*$ is any product either $(\wedge),$ $(\cdot),$ $(\lrcorner,\llcorner)
$ or $($\emph{Clifford product}$).$

In order to prove this property we should use eq.(\ref{SGO.6}) and the
noticeable multivector identity $B\times(X*Y)=(B\times X)*Y+X*(B\times Y),$
with $B\in\bigwedge^{2}V$ and $X,Y\in\bigwedge V.$ By taking into account
eq.(\ref{SGO.3a}) we can see that the following property for the euclidean
scalar product of multivectors holds
\begin{equation}
\underset{\thicksim}{t}_{-}(X)\cdot Y+X\cdot\underset{\thicksim}{t}%
_{-}(Y)=0.\label{SGO.7a}%
\end{equation}
It is consistent with the well-known property: \emph{the adjoint of a
skew-symmetric extensor equals minus the extensor!}

\section{Determinant}

We now define a characteristic scalar associated to any $(1,1)$-extensor $t.$
It is the unique real number, denoted by $\det[t],$ such that
\begin{equation}
\underline{t}(I)=\det[t]I,\label{D.1}%
\end{equation}
for all non-zero pseudoscalar $I.$ It will be called the \emph{determinant of
}$t.$

It is a well-defined\emph{\ }scalar invariant since it does not depend on the
choice of $I.$

We present now some of the most important properties satisfied by the determinant.

\textbf{d1} Let $t$ and $u$ be two $(1,1)$--extensors. It holds
\begin{equation}
\det[u\circ t]=\det[u]\det[t].\label{D.2}%
\end{equation}

Take a non-zero pseudoscalar $I\in\bigwedge^{n}V.$ Then, by using
eq.(\ref{EO.4}) and eq.(\ref{D.1}) we can write that
\begin{align*}
\det[u\circ t]I  &  =\underline{u\circ t}(I)=\underline{u}\circ\underline
{t}(I)=\underline{u}(\underline{t}(I))\\
&  =\underline{u}(\det[t]I)=\det[t]\underline{u}(I),\\
&  =\det[t]\det[u]I.
\end{align*}

\textbf{d2} Let us take $t\in ext_{1}^{1}(V)$ with inverse $t^{-1}\in
ext_{1}^{1}(V).$ It holds
\begin{equation}
\det[t^{-1}]=(\det[t])^{-1}.\label{D.3}%
\end{equation}

Indeed, by using eq.(\ref{D.2}) and the obvious property $\det[i_{V}]=1,$ we
have that
\[
t^{-1}\circ t=t\circ t^{-1}=i_{V}\Rightarrow\det[t^{-1}]\det[t]=\det
[t]\det[t^{-1}]=1.
\]
It means that the \emph{determinant of the inverse equals the inverse of the
determinant.}

Due to the above corollary it is often convenient to use the short notation
$\left.  \det\right.  ^{-1}[t]$ for both of $\det[t^{-1}]$ and $(\det
[t])^{-1}. $

\textbf{d3} Let us take $t\in ext_{1}^{1}(V).$ It holds
\begin{equation}
\det[t^{\dagger}]=\det[t].\label{D.4}%
\end{equation}
Indeed, take a non-zero $I\in\bigwedge^{n}V.$ Then, by using eq.(\ref{D.1}),
eq.(\ref{SAO.3}) and eq.(\ref{SAO.6}), we have that
\[
\det[t^{\dagger}]I\cdot I=\underline{t}^{\dagger}(I)\cdot I=I\cdot
\underline{t}(I)=I\cdot\det[t]I=\det[t]I\cdot I,
\]
from where the expected result follows.

Let $\{e_{j}\}$ be any basis for $V,$ and $\{e^{j}\}$ be its euclidean
reciprocal basis for $V,$ i.e., $e_{j}\cdot e^{k}=\delta_{j}^{k}.$ There are
two interesting and useful formulas for calculating $\det[t],$ i.e.,
\begin{align}
\det[t]  &  =\underline{t}(e_{1}\wedge\ldots\wedge e_{n})\cdot(e^{1}%
\wedge\ldots\wedge e^{n}),\label{D.5a}\\
&  =\underline{t}(e^{1}\wedge\ldots\wedge e^{n})\cdot(e_{1}\wedge\ldots\wedge
e_{n}).\label{D.5b}%
\end{align}
They follow from eq.(\ref{D.1}) by using $(e_{1}\wedge\ldots\wedge e_{n}%
)\cdot(e^{1}\wedge\ldots\wedge e^{n})=1$ which is an immediate consequence of
the formula for the euclidean scalar product of simple $k$-vectors and the
reciprocity property of $\{e_{k}\}$ and $\{e^{k}\}$.

Each of eq.(\ref{D.5a}) and eq.(\ref{D.5b}) is completely equivalent to the
definition of determinant given by eq.(\ref{D.1}).

\section{Some Applications}

\begin{theorem}
Let $(\{b_{k}\},\{b^{k}\})$ and $(\{e_{k}\},\{e^{k}\})$ be two pairs of
euclidean bases for $V.$ There exists an unique invertible $(1,1)$-extensor
$f$ such that
\begin{align}
e_{k}  &  =f(b_{k}),\label{2.41a}\\
e^{k}  &  =f^{\ast}(b^{k})\text{ for each }k=1,\ldots,n.\label{2.41b}%
\end{align}
And, reciprocally given an arbitrary invertible $(1,1)$-extensor, say $f$, it
is possible construct from a pair of reciprocal bases, say $(\{b_{k}%
\},\{b^{k}\}),$ with the above formulas another pair of reciprocal bases, say
$(\{e_{k}\},\{e^{k}\}).$
\end{theorem}

\begin{proof}
Since each one of the sets $\{e_{k}\}$ and $\{e^{k}\}$ is a basis of $V,$
there must be exactly two invertible $(1,1)$-extensors over $V,$ say $f_{1}$
and $f_{2},$ such that
\begin{align*}
e_{k}  &  =f_{1}(b_{k}),\\
e^{k}  &  =f_{2}(b^{k})\text{ for each }k=1,\ldots,n.
\end{align*}
It is not difficult to see that $f_{1}$ and $f_{2}$ are given by
\begin{align*}
f_{1}(v)  &  =\overset{n}{\underset{j=1}{\sum}}(b_{j}\cdot v)e_{j},\\
f_{2}(v)  &  =\overset{n}{\underset{j=1}{\sum}}(b^{j}\cdot v)e^{j}.
\end{align*}
Due to the reciprocity property of $(\{e_{k}\},\{e^{k}\})$ we have
\[
f_{1}(b_{k})\cdot f_{2}(b^{l})=\delta_{k}^{l}\Rightarrow b_{k}\cdot
f_{1}^{\dagger}\circ f_{2}(b^{l})=\delta_{k}^{l}\Rightarrow f_{1}^{\dagger
}\circ f_{2}(b^{l})=b^{l},
\]
for each $l=1,\ldots,n.$ Thus, $f_{1}^{\dagger}\circ f_{2}=i_{V}.$
Now, let us choose $f_{1}=f.$ Then, it must be $f_{2}=f^{*}$ (recall that
$f^{*}=(f^{\dagger})^{-1}=(f^{-1})^{\dagger}$), and so the first statement follows.
To prove the second statement we must check that $\{e_{k}\}$ and $\{e^{k}\}$
given by eqs.(\ref{2.41a}) and (\ref{2.41b}) satisfy the reciprocity property.
Indeed, by using eq.(\ref{SAO.3}), we can write
\[
e_{k}\cdot e^{l}=f(b_{k})\cdot f^{*}(b^{l})=b_{k}\cdot
f^{\dagger}\circ f^{*}(b^{l})=b_{k}\cdot b^{l}=\delta_{k}^{l}.
\]
\end{proof}

\subsection{Orthonormal Bases}

It should be noted that if $f$ is an \emph{orthogonal} $(1,1)$-\emph{extensor}
(i.e., $f^{\dagger}=f^{-1},$ or equivalently $f=f^{*}$ (the adjoint of $f$
being taken with respect to the euclidean scalar product), then $\{e_{k}\}$,
as defined in eq.(\ref{2.41a}), is an \emph{orthonormal basis} for $V,$ i.e.,
$e_{j}\cdot e_{k}=\delta_{jk}$, if and only if $\{b_{k}\}$ is an
\emph{orthonormal basis} for $V,$ i.e., $b_{j}\cdot b_{k}=\delta_{jk}.$
Indeed, $e_{j}\cdot e_{k}=f(b_{j})\cdot f(b_{k})=f^{\dagger}\circ
f(b_{j})\cdot b_{k}=b_{j}\cdot b_{k}=\delta_{jk}$.

\subsection{Changing Basis Extensor}

Theorem 1 implies that for two arbitrary pairs of reciprocal bases of $V,$ say
$(\{e_{k}\},\{e^{k}\})$ and $(\{e_{k}^{\prime}\},\{e^{k\prime}\})$, there must
be an unique invertible $(1,1)$-extensor over $V,$ say $\varepsilon,$ such
that
\begin{align}
\varepsilon(e_{k})  &  =e_{k}^{\prime},\label{2.42a}\\
\varepsilon^{*}(e^{k})  &  =e^{k\prime}.\label{2.42b}%
\end{align}

Indeed, there are exactly two invertible $(1,1)$-extensors, say $f$ and
$f^{\prime},$ such that $e_{k}=f(b_{k}),$ $e^{k}=f^{*}(b^{k})$ and
$e_{k}^{\prime}=f^{\prime}(b_{k}),$ $e^{k\prime}=f^{\prime*}(b^{k})$ for each
$k=1,\ldots,n.$ From these equations we get $e_{k}^{\prime}=f^{\prime}\circ
f^{-1}(e_{k})$ and $e^{k\prime}=f^{\prime*}\circ f^{\dagger}(e^{k}) $. It
means that there is an unique invertible $(1,1)$-extensor which satisfies
eqs.(\ref{2.42a}) and (\ref{2.42b}). Such one is given by $\varepsilon
=f^{\prime}\circ f^{-1}.$

Such $\varepsilon\in ext_{1}^{1}(V)$ will be called the \emph{changing basis
extensor relative to }$(\{e_{k}\},\{e^{k}\})$\emph{\ and }$(\{e_{k}^{\prime
}\},\{e^{k\prime}\})$ (in this order!).

The changing basis extensor $\varepsilon,$ as is not difficult to see, can be
defined equivalently by
\begin{equation}
\varepsilon(v)=(e^{s}\cdot v)e_{s}^{\prime}.\label{2.42c}%
\end{equation}

Also, we can easily see that $\varepsilon^{-1},$ can be alternatively defined
by
\begin{equation}
\varepsilon^{-1}(v)=(e^{s\prime}\cdot v)e_{s},\label{2.42d}%
\end{equation}
and, by using eq.(\ref{SAO.3}), a straightforward calculation yields
\begin{align}
\varepsilon^{\dagger}(v)  &  =(e_{s}^{\prime}\cdot v)e^{s},\label{2.42e}\\
\varepsilon^{*}(v)  &  =(e_{s}\cdot v)e^{s\prime}.\label{2.42f}%
\end{align}

As we know, the vector bases $\{e_{k}\}$ and $\{e_{k}^{\prime}\}$ induce the
$k$-vector bases $\{e_{j_{1}}\wedge\ldots\wedge e_{j_{k}}\}$ and $\{e_{j_{1}%
}^{\prime}\wedge\ldots\wedge e_{j_{k}}^{\prime}\}$. From eq.(\ref{2.42a}) by
using eq.(\ref{EO.2c}) it follows that
\begin{equation}
\underline{\varepsilon}(e_{j_{1}}\wedge\ldots\wedge e_{j_{k}})=e_{j_{1}%
}^{\prime}\wedge\ldots\wedge e_{j_{k}}^{\prime}.\label{2.42g}%
\end{equation}

Analogously for the vector bases $\{e^{j}\}$ and $\{e^{j\prime}\}$ it holds
\begin{equation}
\underline{\varepsilon}^{*}(e^{j_{1}}\wedge\ldots\wedge e^{j_{k}}%
)=e^{j_{1}\prime}\wedge\ldots\wedge e^{j_{k}\prime}.\label{2.42h}%
\end{equation}

\subsection{Inversion of a Non-singular $(1,1)$-Extensor}

We will end this section presenting an useful formula for the inversion of a
non-singular $(1,1)$--extensor.

Let us take $t\in ext_{1}^{1}(V)$. If $t$ is non-singular, i.e., $\det
[t]\neq0,$ then there exists its inverse $t^{-1}\in ext_{1}^{1}(V)$ which is
given by
\begin{equation}
t^{-1}(v)=\left.  \det\right.  ^{-1}[t]\underline{t}^{\dagger}(vI)I^{-1}%
,\label{D.6}%
\end{equation}
where $I\in\bigwedge^{n}V$ is any non-zero pseudoscalar.

\begin{proof}
We must prove that $t^{-1}$ given by the formula above satisfies both of
conditions $t^{-1}\circ t=i_{V}$ and $t\circ t^{-1}=i_{V}.$
Let $I\in\bigwedge^{n}V$ be a non-zero pseudoscalar. Take $v\in V,$ by using
the extensor identities\footnote{These extensor identities follow directly
from the fundamental identity $X\lrcorner\underline{t}(Y)=\underline
{t}(\underline{t}^{\dagger}(X)\lrcorner Y)$ with $X,Y\in\bigwedge V$. For the
first one: take $X=v,$ $Y=I$ and use $(t^{\dagger})^{\dagger}=t,$
eq.(\ref{D.1}) and $\det[t^{\dagger}]=\det[t].$ For the second one: take
$X=vI,Y=I^{-1}$ and use eq.(\ref{D.1}).} $\underline{t}^{\dagger}%
(t(v)I)I^{-1}=t(\underline{t}^{\dagger}(vI)I^{-1})=\det[t]v,$ we have that
\[
t^{-1}\circ t(v)=t^{-1}(t(v))=\left.  \det\right.  ^{-1}[t]\underline
{t}^{\dagger}(t(v)I)I^{-1}=\left.  \det\right.  ^{-1}[t]\det[t]v=i_{V}(v).
\]
And
\[
t\circ t^{-1}(v)=t(t^{-1}(v))=\left.  \det\right.  ^{-1}[t]t(\underline
{t}^{\dagger}(vI)I^{-1})=\left.  \det\right.  ^{-1}[t]\det[t]v=i_{V}(v).
\]
\end{proof}

\section{Conclusions}

We introduced and developed some aspects of the theory of extensors, and made
preliminary applications of it. The concept of extensor when used together
with the euclidean Clifford algebra (as introduced in \cite{4}, paper I of
this series) permits an intrinsic formulation of the key concepts of linear
algebra theory, and plays a crucial role in our study of more sophisticated
concepts which are developed in subsequent papers of the present series. And
also in some forthcoming new series of papers.\emph{\vspace{0.1in}}

\textbf{Acknowledgement: }V. V. Fern\'{a}ndez is grateful to FAPESP for a
posdoctoral fellowship. W. A. Rodrigues Jr. is grateful to CNPq for a senior
research fellowship (contract 201560/82-8) and to the Department of
Mathematics of the University of Liverpool for the hospitality. Authors are
also grateful to Drs. P. Lounesto, I. Porteous, and J. Vaz Jr. for their
interest on our research and for useful suggestions and discussions.

\end{document}